\documentstyle[A4]{article}

\makeatletter

\newfam\mibfam
\font\fivmib=cmmib5
\font\sixmib=cmmib6
\font\svnmib=cmmib7
\font\egtmib=cmmib8
\font\ninmib=cmmib9
\font\tenmib=cmmib10
\font\elvmib=cmmib10 scaled \magstephalf
\font\twlmib=cmmib10 scaled \magstep 1
\font\frtnmib=cmmib10 scaled \magstep 2
\font\svtnmib=cmmib10 scaled \magstep 3
\font\twtymib=cmmib10 scaled \magstep 4
\font\twfvmib=cmmib10 scaled \magstep 5

\@addfontinfo\@vpt{\def\mib{\fam\mibfam\fivmib}
 \textfont\mibfam=\fivmib
 \scriptfont\mibfam=\fivmib
 \scriptscriptfont\mibfam=\fivmib}

\@addfontinfo\@vipt{\def\mib{\fam\mibfam\sixmib}
 \textfont\mibfam=\sixmib
 \scriptfont\mibfam=\sixmib
 \scriptscriptfont\mibfam=\sixmib}

\@addfontinfo\@viipt{\def\mib{\fam\mibfam\svnmib}
 \textfont\mibfam=\svnmib
 \scriptfont\mibfam=\fivmib
 \scriptscriptfont\mibfam=\fivmib}

\@addfontinfo\@viiipt{\def\mib{\fam\mibfam\egtmib}
 \textfont\mibfam=\egtmib
 \scriptfont\mibfam=\sixmib
 \scriptscriptfont\mibfam=\fivmib}

\@addfontinfo\@ixpt{\def\mib{\fam\mibfam\ninmib}
 \textfont\mibfam=\ninmib
 \scriptfont\mibfam=\sixmib
 \scriptscriptfont\mibfam=\fivmib}

\@addfontinfo\@xpt{\def\mib{\fam\mibfam\tenmib}
 \textfont\mibfam=\tenmib
 \scriptfont\mibfam=\svnmib
 \scriptscriptfont\mibfam=\fivmib}

\@addfontinfo\@xipt{\def\mib{\fam\mibfam\elvmib}
 \textfont\mibfam=\elvmib
 \scriptfont\mibfam=\egtmib
 \scriptscriptfont\mibfam=\sixmib}

\@addfontinfo\@xiipt{\def\mib{\fam\mibfam\twlmib}
 \textfont\mibfam=\twlmib
 \scriptfont\mibfam=\egtmib
 \scriptscriptfont\mibfam=\sixmib}

\@addfontinfo\@xivpt{\def\mib{\fam\mibfam\frtnmib}
 \textfont\mibfam=\frtnmib
 \scriptfont\mibfam=\tenmib
 \scriptscriptfont\mibfam=\svnmib}

\@addfontinfo\@xviipt{\def\mib{\fam\mibfam\svtnmib}
 \textfont\mibfam=\svtnmib
 \scriptfont\mibfam=\twlmib
 \scriptscriptfont\mibfam=\tenmib}

\@addfontinfo\@xxpt{\def\mib{\fam\mibfam\twtymib}
 \textfont\mibfam=\twtymib
 \scriptfont\mibfam=\frtnmib
 \scriptscriptfont\mibfam=\twlmib}

\@addfontinfo\@xxvpt{\def\mib{\fam\mibfam\twfvmib}
 \textfont\mibfam=\twfvmib
 \scriptfont\mibfam=\twtymib
 \scriptscriptfont\mibfam=\svtnmib}

\makeatother

\def\a{a\llap{\smash{\raisebox{-1.5ex}{`}}}}

\def\im{{\rm i}}
\def\d#1{\,{\rm d}#1\,}

\renewcommand{\vec}[1]{{\mib#1}}

\newtheorem{assumption}{Assumption}

\newenvironment{braceqns}[1]{
 \renewcommand{\arraystretch}{1.5}
 \begin{equation}
 \left.\begin{array}{#1}}{
 \end{array}\right\}
 \end{equation}
 \renewcommand{\arraystretch}{1}}

\title{\marginpar{\vspace{-1in}\hspace{-1in}\small KFT U{\L} 7/93}Quantum
braided Poincar\'e group\thanks{Supported by KBN grant No. 2 0218 91 01}}
\author{Jakub Rembieli\'nski\thanks{e-mail:
{\tt jaremb@plunlo51.bitnet}}\\[\baselineskip]
Department of Theoretical Physics\\
University of \L\'od\'z\\
ul.~Pomorska 149/153\\
90--236 \L\'od\'z,
Poland}
\date{August 1993}

\begin{document}
\bibliographystyle{unsrt}
\maketitle

\begin{abstract}
A new deformation of the of the Poincar\'e group and of the Minkowski
space-time is given. From the mathematical point of view this deformation is
rather quantum-braided group. Global and local structure of this
quantum-braided Poincar\'e group is investigated. A kind of ``quantum metrics''
is introduced in the $q$-Minkowski space.
\end{abstract}

\section{Introduction}
The idea of quantization of space-time and of the Poincar\'e group by means of
quantum group theory methods was egzamined recently in a number of papers
\cite{kin,poinc,qsymm,wor,lukierski,luk,wess,aref}.  In particular
two-dimensional deformed space-time groups was considered in
\cite{kin,poinc,qsymm}, while the four-dimensional ones by Podle\'s \&
Woronowicz \cite{wor}, Lukierski, Nowicki, Ruegg \& Tolstoy
\cite{lukierski,luk}, Schmidke, Wess \& Zumino \cite{wess}.

In this paper we introduce a new quantization of the four-dimensional Minkowski
space-time and Poincar\'e group.  This deformation has many nice properties,
like isotropy, existence of a kind of   ``quantum geometry'' and relatively
clear physical interpretation.  From the mathematical point of view the
resulting group is rather a quantum-braided group--a hybryde of the both
quantum and braided groups.  Quantum braided group structure was initiated in
\cite{majid,hlavaty}.

The main assumption we made is the following:
\begin{assumption}
The $q$-space-time and $q$-Poincar\'e algebras are isotropic, i.e.\ the
rotation group is an authomorphism group of the above algebras\/. {\em Strictly
speaking we assume that there are no distinguished direction in the space
sector of the space -time.}
\end{assumption}
It seems that this assumption is very strongly supported by experimantal data.
Moreover, as was shown by Bacry and Levy-Leblond \cite{bacry}, rotational
symmetry very strongly restricts possible space-time groups. Notice that
isotropy condition is not fulfiled in the approach presented in the
\cite{wess}.

Under the above assumption the space-time generators $x^\mu$
satisfy
\begin{braceqns}{rcl}
x^ix^j&=&g(x^0,\vec x^2)x^jx^i,\\
x^0\vec x&=&\vec xf(x^0,\vec x^2).\label{wstep}
\end{braceqns}
Taking into account the permutational symmetry we immediately obtain that the
function $g(x^0,\vec x^2)=1$.

In the following we restrict ourselves to an inhomogenous Manin's reordering
rules for every space-time plane, i.e.\ we accept the {\em Ansatz\/}
\begin{equation}
f(x^0,\vec x^2)=ax^0+b.\label{ansatz}
\end{equation}
Now, by means of the hermicity of coordinates
\begin{equation}
{x^\mu}^*=x^\mu,\label{hermicity}
\end{equation}
where $^*$ denotes an antilinear anti-involution in our algebra, we deduce that
eq.~(\ref{wstep}) takes the form
\begin{braceqns}{rcl}x^ix^j&=&x^jx^i,\\
\vec xx^0&=&qx^0\vec x+\im\kappa q^\frac12\vec x\label{kmink}
\end{braceqns}
with $|q|=1$, $\bar\kappa=\kappa$.

Notice that the variant with $q=1$ is a dual form of the
Lukierski-Nowicki-Ruegg $\kappa$ deformation \cite{lukierski,luk,zakrz}.  In
this paper we consider complementary case of an arbitrary phase $q$, $|q|=1$
and $\kappa=0$.  Our choice follows naturally from the requirement of the
invariance under geometrical time reflection; namely (\ref{kmink}) is invariant
under $x^0\to-x^0$, $\vec x\to\vec x$ only for $\kappa=0$. (We know that
$T$-invariance is broken only for a restricted class of interactions, so it
should be connected with dynamics rather than with the space-time algebra).

Summarizing, our reordering rules for space-time generators have the form
\begin{braceqns}{rcl}x^ix^j&=&x^jx^i,\\
x^0\vec x&=&q\vec xx^0,\label{mink}
\end{braceqns}
with $|q|=1$.

\section{Quantum-braided inhomogenous linear group ($QBIGL(4{\mib R})$)}
Now, the (braided) co-module action of the real inhomogenous linear quantum
group is defined as
\begin{equation}
\left(\begin{array}{c}{x^0}'\\ \vec x'\\1\end{array}\right)
=\left(\begin{array}{c|c}\Lambda&\matrix{a^0\cr\vec a}\\
 \hline0&1\end{array}\right)
\left(\begin{array}{c}x^0\\ \vec x\\1\end{array}\right)\label{action}
\end{equation}
with
\begin{equation}
\Lambda=[{{\mit\Lambda}^\mu}_\nu]
\equiv\pmatrix{\beta&q^{-\frac12}\vec u^\dagger\cr q^\frac12\vec w&B}.
\label{lorentz}
\end{equation}
Here ``$^\dagger$' denotes matrix transposition accompaniated with $^*$
conjugation and $B=[{B^i}_j]$.  The generators ${{\mit\Lambda}^\mu}_\nu$ and
$a^\mu$ are real, i.e.\
\begin{equation}
({{\mit\Lambda}^\mu}_\nu)^*={{\mit\Lambda}^\mu}_\nu,\quad
{a^\mu}^*=a^\mu.\label{real}
\end{equation}
As was announced we admit a braiding in our quantum group action.  This means
that $x^\mu$'s do not commute with the group generators
${{\mit\Lambda}^\mu}_\nu$ and $a^\mu$ in general.  Moreover generators of the
next group action ${{{\mit\Lambda}'}^\mu}_\nu$ and ${a'}^\mu$ do not commute
with ${{\mit\Lambda}^\mu}_\nu$ and $a^\mu$ in general.  We will assume the
group composition (co-product) in the form
\begin{equation}
\left(\begin{array}{c|c}\Lambda''&a''\\ \hline0&1\end{array}\right)
=\left(\begin{array}{c|c}\Lambda'&a'\\ \hline0&1\end{array}\right)
\left(\begin{array}{c|c}\Lambda&a\\ \hline0&1\end{array}\right).\label{coprod}
\end{equation}
In the formulas (\ref{action}) and (\ref{coprod}) we do not use the tensor
product notation because of braiding and explicit distinguishing between
$(\Lambda,a)$ and $(\Lambda',a')$.  Now, because in principle we can apply
firstly $(\Lambda',a')$ and next $(\Lambda,a)$ in the eq.~(\ref{coprod}) (with
a different result $(\tilde\Lambda'',\tilde a'')$) then
$({{{\mit\Lambda}'}^\mu}_\nu,{a'}^\mu)$ should satisfy the same reordering
rules with $x^\mu$ as $({{\mit\Lambda}^\mu}_\nu,a^\mu)$. Moreover the
reordering rules between generators $({{\mit\Lambda}^\mu}_\nu,a^\mu)$ and
$({{{\mit\Lambda}'}^\mu}_\nu,{a'}^\mu)$ should be symmetric under intertwining
of primed and unprimed generators.  Taking into account the above remarks and
by means of the isotropy condition we obtain (under the generalized {\em Bethe
Ansatz\/} assumption) the following algebra for generators
\begin{equation}
{{\mit\Lambda}^\mu}_\nu{{\mit\Lambda}^\sigma}_\lambda
={{\mit\Lambda}^\sigma}_\lambda{{\mit\Lambda}^\mu}_\nu, \label{reord1}
\end{equation}
\begin{braceqns}{rclcrcl}
a^0\beta&=&\beta a^0,&\quad&\vec a\beta&=&\beta\vec a,\\
a^0{B^i}_j&=&{B^i}_ja^0,&\quad&\vec a{B^i}_j&=&{B^i}_j\vec a,\\
a^0\vec w&=&q\vec wa^0,&\quad&\vec aw^i&=&qw^i\vec a,\\
a^0\vec u&=&q^{-1}\vec ua^0,&\quad&\vec au^i&=&q^{-1}u^i\vec a,\label{reord2}
\end{braceqns}
\begin{equation}
a^0\vec a=q\vec aa^0,\qquad a^ia^j=a^ja^i,\label{reord3}
\end{equation}
The co-module braiding rules takes the form
\begin{braceqns}{rclcrcl}
x^0\beta&=&\beta x^0,&\quad&\vec x\beta&=&\beta\vec x,\\
x^0{B^i}_j&=&{B^i}_jx^0,&\quad&\vec x{B^i}_j&=&{B^i}_j\vec x,\\
x^0\vec w&=&q\vec wx^0,&\quad&\vec xw^i&=&qw^i\vec x,\\
x^0\vec u&=&q^{-1}\vec ux^0,&\quad&\vec xu^i&=&q^{-1}u^i\vec x,\\
x^0a^0&=&a^0x^0,&\quad&\vec xa^0&=&q^{-1}a^0\vec x,\\
x^0\vec a&=&q\vec ax^0,&\quad&\vec xa^i&=&a^i\vec x.
\label{reord4}
\end{braceqns}
The same relations hold for ${{{\mit\Lambda}'}^\mu}_\nu$ and ${a'}^\mu$.
The braiding rules reads
\begin{equation}
{{\mit\Lambda}^\mu}_\nu{{{\mit\Lambda}'}^\sigma}_\lambda
={{{\mit\Lambda}'}^\sigma}_\lambda{{\mit\Lambda}^\mu}_\nu, \label{reord5}
\end{equation}
\begin{braceqns}{rclcrcl}
a^0\beta'&=&\beta'a^0,&\quad&\vec a\beta'&=&\beta'\vec a,\\
a^0{{B'}^i}_j&=&{{B'}^i}_ja^0,&\quad&\vec a{{B'}^i}_j&=&{{B'}^i}_j\vec a,\\
a^0\vec w'&=&q\vec w'a^0,&\quad&\vec a{w'}^i&=&q{w'}^i\vec a,\\
a^0\vec u'&=&q^{-1}\vec u'a^0,&\quad&\vec a{u'}^i&=&q^{-1}{u'}^i\vec a,
\label{reord6}
\end{braceqns}
\begin{braceqns}{rclcrcl}
{a'}^0\beta&=&\beta{a'}^0,&\quad&\vec a'\beta&=&\beta\vec a',\\
{a'}^0{B^i}_j&=&{B^i}_j{a'}^0,&\quad&\vec a'{B^i}_j&=&{B^i}_j\vec a',\\
{a'}^0\vec w&=&q\vec w{a'}^0,&\quad&\vec a'w^i&=&qw^i\vec a',\\
{a'}^0\vec u&=&q^{-1}\vec u{a'}^0,&\quad&\vec a'u^i&=&q^{-1}u^i\vec a',
\label{reord7}
\end{braceqns}
\begin{braceqns}{rclcrcl}
a^0{a'}^0&=&{a'}^0a^0,&\quad&\vec a{a'}^0&=&q^{-1}{a'}^0\vec a,\\
a^0\vec a'&=&q\vec a'a^0,&\quad&a^i{a'}^j&=&{a'}^ja^i,
\label{reord8}
\end{braceqns}
Co-unity is defined standardly as
\begin{braceqns}{rcccl}
\epsilon(\Lambda)&=&\epsilon(\Lambda')&=&I,\\
\epsilon(a^\mu)&=&\epsilon({a'}^\mu)&=&0,
\label{counit}
\end{braceqns}
where $I$ is the $4\times4$ unit matrix.  The antipode has the form
\begin{equation}
\left(\begin{array}{c|c}\Lambda^{-1}&-\Lambda^{-1}a\\
\hline0&1\end{array}\right)\label{antipode}
\end{equation}
where $\Lambda^{-1}$ is understand in the usual sense as a
matrix inverse because $\{{{\mit\Lambda}^\mu}_\nu\}$ is a {\em
commutative\/} subalgebra of our quantum braided group (see
eqs.~(\ref{reord1}), (\ref{reord5})).  In particular
$\det\Lambda$ is of the standard form and belongs to the center
of the above algebra.

\section{Quantum Minkowski space-time and the quantum braided Poincar\'e group}
To select in $QBIGL(4{\mib R})$ an appriopriate Poincar\'e subgroup it is
necessary to define a substitute of geometry in our space-time algebra.  To
do this let us firstly introduce a covariant differential calculus in that
algebra.   Taking into account our isotropy assumption, the
algebra (\ref{mink}) and the group action (\ref{action}), we
obtain, with help of the classification \cite{holomorph} of the
differential calculi, the following covariant reordering rules
for differentials
\begin{braceqns}{rcl}
x^0\d{x^0}\!&=&\!\d{x^0}x^0,\\
x^i\d{x^k}\!&=&\!\d{x^k}x^i,\\
x^0\d{\vec x}\!&=&q\d{\vec x}x^0,\\
\vec x\d{x^0}\!&=&q^{-1}\d{x^0}\vec{x},
\label{diff1}
\end{braceqns}
\begin{braceqns}{rcl}
(\!\d{x^0}\!)^2&=&0,\\
\!\d{x^i}\!\d{x^j}\!&=&-\!\d{x^j}\!\d{x^i},\\
\!\d{x^0}\!\d{\vec x}\!&=&-q\d{\vec x}\!\d{x^0}.
\label{diff2}
\end{braceqns}
Now, a trajectory in the bundle of the space-time algebras is given by a
parametric dependence of $x^\mu$ on an affine parameter, say $\tau$, i.e.\
\begin{equation}
x^\mu=x^\mu(\tau),\label{tau}
\end{equation}
so
\begin{equation}
\!\d{x^\mu}\!=\dot x^\mu(\tau)\d{\tau}\!,\label{vel}
\end{equation}
where the ``four-velocity'' $\dot x^\mu$ satisfy, according to the
eq.~(\ref{diff1}) and the covariance condition, the following
first order differential calculus rules
\begin{braceqns}{rcl}
x^0\dot x^0&=&\dot x^0x^0,\\
x^i\dot x^k&=&\dot x^kx^i,\\
x^0\dot x^k&=&q\dot x^kx^0,\\
x^k\dot x^0&=&q^{-1}\dot x^kx^0,\\
\dot x^i\dot x^j&=&\dot x^j\dot x^i,\\
\dot x^0\dot x^i&=&q\dot x^i\dot x^0.
\label{dot}
\end{braceqns}
Now, we try to find a substitute of the relativistic line element
\begin{equation}
\!\d{s^2}\!=({\dot x^0{}}^2-{\dot{\vec x}}^2)\d{\tau^2}\!.\label{line}
\end{equation}
We see that $\!\d{s^2}\!$ in the above form does not belong to the center of
our algebra.  Therefore we have difficulties with interpretation of $\!\d{s}\!$
as a line parameter and consequently with a formulation of the action
principle, definition of geodesics, free motion, etc.  To omit this difficulty
let us introduce a ``quantum metric'' $g_{\mu\nu}$
\cite{poinc,sitarz} satisfying reality and isotropy condition; namely we
define
\begin{equation}
\!\d{s^2_q}\!=\dot x^\mu g_{\mu\nu}\dot x^\nu\d{\tau^2}\!
=(\dot x^\dagger g\dot x)\d{\tau^2}\!,\label{qline}
\end{equation}
with covariant, ``quantum flat'' metric tensor
\begin{equation}
g=[g_{\mu\nu}]
=\pmatrix{\gamma&0&0&0\cr0&-\Gamma&0&0\cr0&0&-\Gamma&0\cr0&0&0&-\Gamma}
\label{g}
\end{equation}
where the new generators are hermitean i.e.\ $\gamma^*=\gamma$,
$\Gamma^*=\Gamma$ and $\dot\gamma=\dot\Gamma=0$.  The square of
the line lement $\!\d{s^2_q}\!$ belongs to the center if the
space-time algebra is completed by
\begin{braceqns}{rcl}
\gamma\Gamma&=&q^4\Gamma\gamma,\\
x^0\gamma&=&\gamma x^0,\\
x^0\Gamma&=&q^{-2}\Gamma,\\
\vec x\gamma&=&q^2\gamma\vec x,\\
\vec x\Gamma&=&\Gamma\vec x.
\label{gamma}
\end{braceqns}
The algebra generated by $x^\mu$ and equipped with the quantum
metrtics $\!\d{s^2_q}\!$ will be called {\em quantum Minkowski
space-time\/}.  Now, we are able to select in $QBIGL(4{\mib R})$
group the quantum braided Poincar\'e group ($QBP$).  It is
defined in analogy to the usual Poincar\'e group as an invariance
group of the line element (\ref{qline}).  Consequently the
generators ${{\mit\Lambda}^\mu}_\nu$ should satisfy the matrix
equation
\begin{equation}
\Lambda^\dagger g\Lambda=g\label{LgL}
\end{equation}
with $g$ defined in the eqs.~(\ref{g}--\ref{gamma}).  By
universality, $\Lambda'$ satisfy also eq.~(\ref{LgL}) so
$\Lambda'\Lambda$ belongs to the same category.

It is necessary to complete the algebra (\ref{gamma}) by
reordering rules consistent with the eq.~(\ref{LgL}); we will
use the parametrization (\ref{lorentz}) of $\Lambda$:
\begin{braceqns}{rclcrclcrcl}
\beta\gamma&=&\gamma\beta,&\quad
&\vec w\gamma&=&q^2\gamma\vec w,&\quad
&a^0\gamma&=&\gamma a^0,\\
\beta\Gamma&=&\Gamma\beta,&\quad
&\vec w\Gamma&=&q^2\Gamma\vec w,&\quad
&a^0\Gamma&=&q^{-2}\Gamma a^0,\\
{B^i}_j\gamma&=&\gamma{B^i}_j,&\quad
&\vec u\gamma&=&q^{-2}\gamma\vec u,&\quad
&\vec a\gamma&=&q^2\gamma\vec a,\\
{B^i}_j\Gamma&=&\Gamma{B^i}_j,&\quad
&\vec u\Gamma&=&q^{-2}\Gamma\vec u,&\quad
&\vec a\Gamma&=&\Gamma\vec a.
\label{GL}
\end{braceqns}
The same relations hold for primed generators ${{{\mit\Lambda}'}^\mu}_\nu$.

Now, we can solve the constraints (\ref{LgL}) by means of the
``polar decomposition'' of $\Lambda$
\renewcommand{\arraystretch}{1}
\begin{equation}
\Lambda=\left(\begin{array}{c|c}\beta&q^{-\frac12}G\vec w\\ \hline
q^\frac12\vec w&I+\frac G{\beta+1}\vec w\times\vec w^\dagger\end{array}\right)
\left(\begin{array}{c|c}1&0\\ \hline 0&R\end{array}\right)
\equiv L(\vec w)R.\label{Lambda}
\end{equation}
Here
\begin{braceqns}{rcl}
{R^i}_k^*&=&{R^i}_k,\\
R^{\rm T}R&=&RR^{\rm T}=I\label{R}
\end{braceqns}
and ${R^i}_k$'s belong to the center of the whole algebra.
Thus ${R^i}_k$ generate the $O(3)$ group.

The new hermitean generator $G$ in the eq.~(\ref{Lambda}) is defined by
\begin{equation}
G=q^2\gamma^{-1}\Gamma\label{G}
\end{equation}
so it satisfy
\begin{braceqns}{rcl}
{{\mit\Lambda}^\mu}_\nu G&=&G{{\mit\Lambda}^\mu}_\nu,\\
a^\mu G&=&q^{-2}Ga^\mu,\\
x^\mu G&=&q^{-2}Gx^\mu.\label{xG}
\end{braceqns}
Furthremore $\beta$ and $\vec w$ fulufil the constraint
\begin{equation}
\beta^2-G\vec w^2=1.\label{constr}
\end{equation}
Notice that apperance of $G$ in the above relations reflects the
fact that the generators ${{\mit\Lambda}^\mu}_\nu$ have mixed (co-
and contra-variant) tensor nature and the ``metric tensor'' is
given by (\ref{g}).  Notice also that inverse of the boost
$L^{-1}(\vec w)=L(-\vec w)$ as in the classical case.

Summarising, $QBP$ group contains translation $a^\mu$ and
$q$-Minkowski rotations ${{\mit\Lambda}^\mu}_\nu$ satisfying
(\ref{LgL}) (or (\ref{constr}) in the parametrization
(\ref{Lambda})) and the corresponding multiplication and
braiding rules hold (eqs.~(\ref{reord1}--\ref{reord8})).

It is interesting to note that contrary to the scheme proposed
in \cite{lukierski,luk,wess} the $QBP$ group has Lorentz group and
translations group as subgroups.

\section{Local structure of $QBP$ group}
An important feature of $QBIGL$ group is that the diagonal
generators $\beta$ and ${B^i}_j$ belong to the center of the
whole group and the space-time algebra, while the off-diagonal
generators $\vec w$, $\vec u$, $a^\mu$ satisfy homogenous
reordering rules. This enables us to apply infinitesimal-like
methods in that case by replacing $\beta\to1+\delta\beta$,
${B^i}_j\to{\delta^i}_j+\delta{B^i}_j$, $\vec w\to\delta\vec w$,
$\vec u\to\delta\vec u$, $a^\mu\to\delta a^\mu$ and under the
same algebraic rules. This is equivalent with use of the
corresponding Cartan-Maurer forms and appriopriate definitions
of commutators according to the reordering rules for
Cartan-Maurer forms. In the case of $QBP$ group both methods
gives us the following algebra for ``infinitesimal generators''
of rotations ($J_i$), boosts ($K_i$) and translations ($P_\mu$):
\begin{braceqns}{rclcrcl}
[J_i,J_j]&=&\im\varepsilon_{ijk}J_k,&\quad
&[K_i,K_j]&=&\!-\im G\varepsilon_{ijk}J_k,\\
{}[J_i,k_j]&=&\im\varepsilon_{ijk}K_j,&\quad
&[J_i,P_0]&=&0,\\
{}[J_i,P_j]&=&\im\varepsilon_{ijk}P_k,&\quad
&[P_i,P_j]&=&0,\\
{}[P_0,P_i]_q&=&0,&\quad
&[P_0,K_i]_q&=&\!-\im q^\frac12P_i,\\
{}[P_i,K_j]_q&=&\!-\im q^\frac32G\delta_{ij}P_0,&&&&\label{lie}
\end{braceqns}
where $[A,B]_q=AB-qBA$, $[A,B]=[A,B]_1$.
The detailed identification of Lie generators and the algebra of
Cartan-Maurer forms is given in the Appendix.

Now, in a unitary realization of $QBP$, $\vec J$, $\vec K$ and
$P_\mu$ are hermitean and by consistency
\begin{braceqns}{rclcrcl}
\vec J\gamma&=&\gamma\vec J,&\quad
\vec K\gamma&=&q^{-2}\gamma\vec K,\\
\vec J\Gamma&=&\Gamma\vec J,&\quad
\vec K\Gamma&=&q^{-2}\Gamma\vec K,\\
P_0\gamma&=&\gamma P_0,&\quad
P_k\gamma&=&q^{-2}\gamma P_k,\\
P_0\Gamma&=&q^2\Gamma P_0,&\quad
P_k\Gamma&=&\Gamma P_k,\label{lie-g}
\end{braceqns}
as well as
\begin{itemize}
\item $\vec J$ and $\vec K$ commute with ${R^i}_j$, $\vec w$, $\beta$ and $G$,
i.e.\ $[J_k,{R^i}_j]=[J_k,\vec w]=[J_k,\beta]=[J_k,G]=0$,
$[K_k,{R^i}_j]=[K_k,\vec w]=[K_k,\beta]=[K_k,G]=0$;
\item $\vec J$ commute with $a^\mu$: $[J_k,a^\mu]=0$;
\item $\vec K$ $q$-commute with $a^\mu$, i.e.\ $[K_i,a^\mu]_q=0$;
\item $P_\mu$ commute with ${R^i}_j$, $\beta$.
\end{itemize}
and
\begin{braceqns}{rclcrcl}
[\vec w,P_\mu]_q&=&0,&\quad&[P_\mu,{R^i}_j]&=&[P_\mu,\beta]=0,\\
a^0P_0&=&P_0a^0,&\quad&\vec aP_0&=&qP_0\vec a,\\
a^kP_i&=&P_ia^k,&\quad&a^0P_i&=&q^{-1}P_ia^0.
\end{braceqns}
Quadratic Casimir $P^2=P_\mu g_{\mu\nu}P_\nu=P_\mu P^\mu$.

It is easy to construct a realization of the algebra
(\ref{lie}) in the space of functions over the space-time algebra:
\begin{braceqns}{rcl}
J_k&=&-\im\varepsilon_{kij}(x^i\partial_j-x^j\partial_i),\\
K_i&=&-\im(q^{-\frac12}Gx^i\partial_0+q^\frac12x^0\partial_i),\\
P_\mu&=&-\im\partial_\mu.
\end{braceqns}
Here $\partial_\mu$ are the corresponding Jackson derivatives
defined via $\!\d{f}\!=\!\d{x^\mu}\partial_\mu f$.

\section{Kinematics}
Let us define the (contravariant) four-momentum $p^\mu$ of a
free particle by the standard formula
\begin{equation}
p^\mu=m\dot x^\mu. \label{mom}
\end{equation}
Now, a consistency of the above definition with the geometric
interpretation of $p_\mu=g_{\mu\nu}p^\nu$ as translation
generators (see eq.~(\ref{lie}) demands commutativity of the
inertial mass $m$, i.e.\ $m$ belongs to the center of the
space-time (and group) algebra.  Moreover, $m$ is hermitean
(i.e.\ $m^*=m$).  By means of the eqs.~(\ref{dot}) and
(\ref{gamma}) we obtain
\begin{braceqns}{rclcrclcrcl}
p^0\vec p&=&q\vec pp^0,&\quad&x^0\vec p&=&q\vec px^0,&\quad&x^0p^0&=&p^0x^0,\\
p^ip^j&=&p^jp^i,&\quad&x^ip^j&=&p^jx^i,&\quad&x^0\vec p&=&q\vec px^0\label{pp}
\end{braceqns}
and
\begin{braceqns}{rclcrcl}
p^0\gamma&=&\gamma p^0,&\quad&\vec p\gamma&=&q^2\gamma\vec p,\\
p^0\Gamma&=&q^{-2}\Gamma p^0,&\quad&\vec p\Gamma&=&\Gamma\vec p.\label{pg}
\end{braceqns}
The braid relations between $p^\mu$ and
$({{\mit\Lambda}^\mu}_\nu,a^\mu)$ can be obtained by means of
the eqs.~(\ref{reord4}).

Notice that
\begin{equation}
p^2=p^\mu g_{\mu\nu} p^\nu=m^2c^2 \label{p2}
\end{equation}
with the square of the light velocity
\begin{equation}
c^2=\dot x^\mu g_{\mu\nu}\dot x^\nu, \label{c2}
\end{equation}
where $\dot c=0$ by means the free motion $\ddot x^\mu=0$ equation.
Therefore the light velocity is a constant of motion.

Free particle trajectories (geodesics) can be obtained via the
least action principle
\begin{equation}
\delta\int\d{s}\!=0, \label{s}
\end{equation}
where $\!\d{s}\!=\sqrt{\dot x^\mu g_{\mu\nu}\dot x^\nu}\d{\tau}\!$.

If we identify two affine parameters $s$ and $\tau$ via
\begin{equation}
\!\d{s}\!=c\d{\tau}\!
\end{equation}
then the least action principle implies
\begin{equation}
\ddot x^\mu=0
\end{equation}
for geodesics. This confirms our claim about constancy of the light velocity.

\section{Conclusions}
In this paper we introduce a quantum-braided variant of the deformed Poincar\'e
group under isotropy assumption. In its structure $QBP$ group is more close to
the standard Poincar\'e group than other deformations. In particulary it admits
a kind of quantum Minkowski geometry.  Also infinitesimal-like methods can be
applied; it is important from the point of view of identification of physical
observables and construction of representations.

It is interesting an analogy of $QBP$ group with supersymmetric groups; namely
the braiding rules in the supersymmetric case lies in the anticommutativity of
group parameters and coordinates.

Many questions are open. For example an open question is physical
identification of the deformation parameter $q$. Furthermore it is clear that
(like in general relativity) we should distinguish between dynamical time and
coordinate time.  The first one can be identified with the invariant parameter
$\tau$ ($s$) and permit us to causal ordering of events. For the other hand
$x^0$ sholud be interpreted as a coordinate time.  We have no direct relation
between $\tau$ and $x^0$ even in the rest frame ($\tau$ cannot be interpreted
as the proper time). It seems that from the point of view of the measurment
process, the directly measured observable are $x^\mu$'s so $x^0$ is measured
directly rather than $\tau$.

A more detailed discussion of these rather delicate questions will be given in
the forthcoming paper.

\section*{Acknowledgements}
I am grateful for interesting discussions to P.~Kosi\'nski, J.~Lukierski,
H.~Ruegg, K.~A.~Smoli\'nski and S.~Zakrzewski.

\appendix
\makeatletter
\@addtoreset{equation}{section}
\makeatother
\renewcommand{\theequation}{\thesection.\arabic{equation}}
\section{Appendix}
The Cartan-Maurer differential form is defined as
\begin{equation}
\Theta=\left(\begin{array}{c|c}\Lambda^{-1}&-\Lambda^{-1}a\\
\hline0&1\end{array}\right)\d{\left(\begin{array}{c|c}\Lambda&a\\
\hline0&1\end{array}\right)}\!
=\left(\begin{array}{c|c}\Lambda^{-1}\d{\Lambda}\!&\Lambda^{-1}\d{a}\!\\
\hline0&0\end{array}\right)
\end{equation}
Now the Lie generators are idntified via
\begin{eqnarray}
\Omega\equiv\Lambda^{-1}\d{\Lambda}\!&=&\im{{\mit\Omega}^\mu}_\nu{J_\mu}^\nu\\
\Lambda^{-1}\d{a}\!&=&\im\rho^\mu P_\mu.
\end{eqnarray}
Putting
\begin{eqnarray}
\phi^k&=&\varepsilon^{kij}{{\mit\Omega}^i}_j,\\
\kappa^k&=&q^{-\frac12}{{\mit\Omega}^k}_0
\end{eqnarray}
we can show that
\begin{equation}
\vec\phi^*=\vec\phi,\quad\vec\kappa^*=\vec\kappa,\quad{\rho^\mu}^*=\rho^\mu
\end{equation}
A bi-covariant differential calculus for $QBP$ is defined by the
following reordering rules
\begin{braceqns}{rcl}
\phi^i\phi^k&=&-\phi^k\phi^i,\\
\vec\phi\rho^\mu&=&-\rho^\mu\vec\phi,\\
\phi^i\kappa^k&=&-\kappa^k\phi^i,\\
\kappa^i\kappa^k&=&-\kappa^k\kappa^i,\\
\kappa^i\rho^\mu&=&-q^{-1}\rho^\mu\kappa^i,\\
\rho^0\vec\rho&=&-q\vec\rho\rho^0,\\
\rho^i\rho^k&=&-\rho^k\rho^i,
\end{braceqns}
\begin{braceqns}{rcl}
\phi^k{{\mit\Lambda}^\mu}_\nu&=&{{\mit\Lambda}^\mu}_\nu\phi^k,\\
\phi^ka^\mu&=&a^\mu\phi^k,\\
\kappa^i{{\mit\Lambda}^\mu}_\nu&=&{{\mit\Lambda}^\mu}_\nu\kappa^i,\\
\kappa^ia^\mu&=&q^{-1}a^\mu\kappa^i,\\
\rho^0a^0&=&a^0\rho^0,\\
\rho^0a^i&=&qa^i\rho^0,\\
\rho^ia^k&=&a^k\rho^i,\\
\rho^ia^0&=&q^{-1}a^0\rho^i.
\end{braceqns}
Furthermore
\begin{braceqns}{rclcrcl}
\phi^i\gamma&=&\gamma\phi^i,&\quad&\phi^i\Gamma&=&\Gamma\phi^i,\\
\kappa^i\gamma&=&q^2\gamma\kappa^i,&\quad&\kappa^i\Gamma&=&q^2\Gamma\kappa^i,\\
\rho^i\gamma&=&q^2\gamma\rho^i,&\quad&\rho^i\Gamma&=&\Gamma\rho^i,\\
\rho^0\gamma&=&\gamma\rho^0,&\quad&\rho^0\Gamma&=&q^{-2}\Gamma\rho^0.
\end{braceqns}
Finally, the generators $J_k$ and $K_i$ are realted to ${J_mu}^\nu$ by
\begin{eqnarray*}
J_k&=&\varepsilon_{kij}{J_i}^j,\\
K_k&=&{J_k}^0+G{J_0}^k.
\end{eqnarray*}


\end{document}